\documentclass{mem}
\usepackage{natbib}\usepackage{txfonts}\usepackage{balance}
\usepackage{graphicx}
\usepackage[a4paper,breaklinks,dvipdfm]{hyperref}
\idline{75}{282}
\begin{document}

\title{
Multi-wavelength observations of H.E.S.S. AGN
}

   \subtitle{}

\author{
M. \,{Tluczykont\inst{1}~{for~the~H.E.S.S.~Collaboration}}\inst{2} 
          }

  \offprints{M. Tluczykont}

\institute{
Institut f\"ur Experimentalphysik,
Universit\"at Hamburg,
Luruper Chaussee 149,
22761, Hamburg, Germany
\and
{\tt http://www.mpi-hd.mpg.de/hfm/HESS/pages/collaboration/}\\
\email{martin.tluczykont@physik.uni-hamburg.de}
}

\authorrunning{Tluczykont }

\titlerunning{H.E.S.S. AGN}

\abstract{
Multi-frequency observations are a powerful tool of astrophysical investigation.
Not only is data in each wavelength band providing different
clues to the objects nature, but taken simultaneously, these data can reveal the mechanisms
at work in astrophysical objects.
In the past years, joint multi-frequency observations with the H.E.S.S. telescopes
in the very high energy (VHE, E$>$100\,GeV) band and several other experiments in the radio, optical, X-ray, and
high energy (HE, E$>$100\,MeV) bands have lead to intriguing results
that will ultimately help answering the open questions of the location of the very high energy
emission, details of the acceleration mechanism, and the role of the central black hole.

\keywords{Galaxies: active -- Gamma rays: observations -- X-rays: galaxies}
}
\maketitle{}

\section{Introduction}
\label{tluczykont_introduction}
Very high energy (VHE, E$>$100\,GeV) gamma-ray astronomy is the domain of Earth-bound air-shower experiments.
The direction, energy and particle type (gamma-hadron separation) can be reconstructed via
the measurement of air-shower particles, or by Cherenkov photons emitted by them.
Imaging air Cherenkov telescopes (IACTs) make an image of the shower in
Cherenkov light.

In the last decade, the IACT technique has made the step from the pioneering era,
dominated by experiments such as Whipple, HEGRA, CAT, or CANGAROO,
to the now ongoing first physics era, with H.E.S.S. \citep{2006A&A...457..899A}, MAGIC \citep{2008ApJ...674.1037A},
and VERITAS \citep{2008ApJ...679.1427A}.
As compared to the non-imaging technique,
advantages of IACTs are their good angular resolution ($<0.1^\circ$), a relative energy resolution around
15\,\%, a low energy threshold (E\,$>$\,100\,GeV for a 100\,m$^2$ mirror dish) and
a very good sensitivity to point-like sources.
Today, Crab Nebula like sources can be detected in the VHE regime in $\sim$30\,s (5\,$\sigma$ significance level).
As compared to the previous generation of IACTs, a significant leap forward has been achieved.
Not only has the number of known Galactic and extragalactic VHE sources sharply
increased, see e.g. \citet[][]{2010arXiv1005.2330G} for an overview,
or \citet[][]{raue:extragalacticsky} for recent H.E.S.S. results.
The quality of the measurements also improved to the point of resolving lightcurves
down to the minute time-scale, spectra extending from 100\,GeV to beyond 10\,TeV,
and sensitivities allowing to detect VHE emission from quiescent AGN (see below).

The H.E.S.S. experiment is an array of four IACTs located in Namibia in the Khomas highlands,
for observations of Galactic and extragalactic sources in the energy range from 100\,GeV to several 10\,TeV. 

AGN are an ideal target for multi-frequency observations.
Their spectral energy distribution (SED)
reaches from radio waves via optical, X-rays, to very-high-energies.
In the widely-accepted unified model
of AGN \citep[e.g.][]{1984ARA&A..22..471R,1995PASP..107..803U},
the ``central engine'' of these objects consists
of a thin accretion disk of gas, accreting onto
a super-massive central black hole (up to 10$^9$\,M$_\odot$),
and surrounded by a dust torus.
In radio-loud AGN, i.e. objects with a radio to B-band flux ratio
F$_\mathrm{5GHz}$/F$_\mathrm{B}$\,$>$\,10,
two relativistic plasma outflows (jets) {presumably} perpendicular to the
plane of the accretion disk have been observed.
AGN are known to be VHE $\gamma$-ray emitters
since the detection of Mrk\,421 above 300\,GeV by the
Whipple group \citep{1992Natur.358..477P}.
Almost all AGN that were detected in the VHE regime belong to the
class of BL Lac objects, i.e. AGN having their jet pointing at
a small angle to the line of sight.
VHE gamma-rays were also detected from two radio galaxies, M\,87 (see below),
and Cen\,A (\citet{2009ApJ...695L..40A}, also see \citet{rieger_cena}, these proceedings),
the starburst galaxy NGC\,253 \citep{2009Sci...326.1080A} and the FSRQs
3C279 \citep{2011A&A...530A...4A} and PKS\,1222+21 \citep{2011ApJ...730L...8A}.

The characteristic two-peak SED of AGN can be explained
by synchrotron emission of relativistic electrons,
and an inverse Compton (IC) peak, resulting from scattering of electrons
off a seed-photon population, see e.g. \citet[][]{2001AIPC..558..275S} and references therein.
Alternatively, hadronic radiation models explain the emission via the interactions
of relativistic protons with matter \citep{pohl:2000a}, ambient photons \citep{mannheim:1993a}
or magnetic field \citep{aharonian:2000c}, or {photons and magnetic field} \citep{muecke:2001a}.
The observed VHE emission shows high {variability}
ranging from short bursts on a minute time-scale,
to long-term activity of the order of months.
Studies of the variability of BL Lac type objects can contribute to the understanding of
their intrinsic acceleration mechanisms \citep[e.g.][]{krawczynski:2001a,aharonian:2002e}.
%
VHE gamma-rays can be absorbed via e$^+$e$^-$ pair production
with the photons of the extragalactic background light (EBL).
Observations of distant objects in the VHE band therfore provide an indirect measurement of the
SED of the EBL, see e.g. \citet{stecker:1992a,primack:1999a} and references therein.
Open questions of VHE emission in AGN persist:
\begin{list}{-}{}
\item
where in the jet does gamma-ray emission occur? While it is believed that the VHE emission must be beamed
and thus originate somewhere along the jet axis, the exact location is still unknown.
\item
What is the source of the soft seed photons for the inverse Compton effect? Seed photon fields
can originate in different regions, such as the accretion disk, different parts of the jet,
the broad line region, the dusty torus, and synchrotron photons. The latter are favoured,
considering the practically feature-less spectra in the optical and ultra-violet, and for
the sake of gamma-ray transparency (local radiation fields must be low enough to avoid intrinsic absorption by pair production).
\item
Finally, what is the role of the central black hole in powering the jet?
The transfer of binding energy of accreting gas to the jet as well as spin-energy extraction via
the Blandford-Znajek mechanism \citep{1977MNRAS.179..433B} are discussed.
\end{list}

\section{Multi-frequency observations}
\subsection{Overview}
A number of AGN were observed by H.E.S.S. since the start of operation in 2002. Some sources were
observed in the frame of organized multi-frequency campaings, or regular monitoring campaigns,
while others were triggered by high-state observations reported by other instruments.
Table~\ref{mwl_agn_table} gives an overview of H.E.S.S. observations of AGN for which
simultaneous multi-frequency data was taken.
\begin{table*}[ht]
\caption{Summary of published simultaneous multi-wavelength data sets of H.E.S.S. AGN.}
\label{mwl_agn_table}
\begin{center}
\begin{tabular}{llll}
\hline
\\
Object		& Years			&MWL					&arXiv:astro-ph/\\
		&			&					&or reference\\
\hline
\\
PKS 2155-304	& 2003,-06,-08		&RXTE, ROTSEIII, NRT, 			&0506593v1,\\
		&			&Bronberg opt. RXTE,			&0906.2002v1,\\
		&			&Chandra, Swift, ATOM			&0903.2924v1\\
H 2356-309	& 2004-2007		&RXTE, ROTSEIII, NRT			&0607569v1,\\
		&			&XMM, ATOM, NRT				&1004.2089v2\\
PKS 2005-489	& 2004-2007		&XMM-Newton, RXTE			&0911.2709v2\\
1ES 1101-232	& 2004,2005		&XMM-Newton, RXTE, optical		&07052946v2\\
Mrk\,421	& 2004			&RXTE					&0506319v1\\
1ES 0347-121	& 2006			&Swift					&0708.3021v1\\
PG 1553+113	& 2005/2006		&VLT (SINFONI)				&0710.5740v2\\
RGB J0152+017	& 2007			&RXTE, Swift, ATOM			&0802.4021v2\\
PKS0548-322	& 2004-2008		&Swift					&1006.5289v2\\
M 87		& 2003-2008		&Chandra, VLBA				&\citet{2006Sci...314.1424A}\\
		&			&					&\citet{2009Sci...325..444A}\\
AP Librae	& 2010			&Swift, RXTE, Fermi			&\citet{fortin:aplib}\\
HESS\,J1943+213	& 2011			&Fermi, NRT				&1103.0763v1\\
\\
\hline
\end{tabular}
\end{center}
\end{table*}
\subsection{H\,2356-309 and 1ES\,1101-232}
\begin{figure*}[t!]
\resizebox{\hsize}{!}{\includegraphics[clip=true]{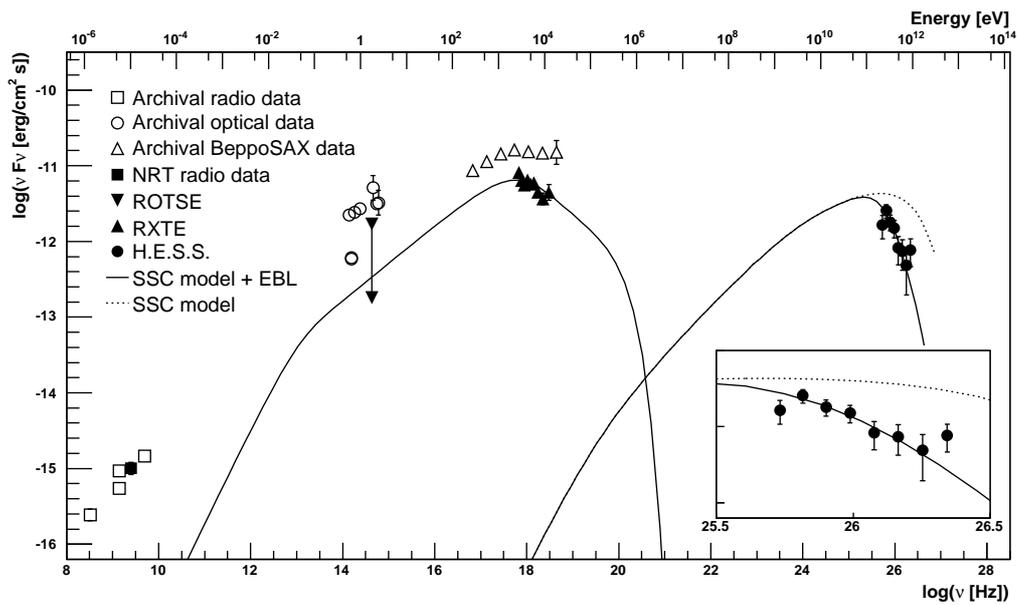}}
\caption{\footnotesize
Historically, this SED of H\,2356$-$309 from 2004 was
the first SED comprising simultaneous radio, optical, X-ray and VHE observations.
                Simultaneous data are shown as filled symbols.
                All other (archival) data are shown as open symbols.
                {Radio emission arises from regions further out in the jet.}
                A single-zone SSC model taking absorption into account is shown as a solid line (dashed line without absorption).
}
\label{h2356_first_sed}
\end{figure*}

The high frequency peaked BL Lac object (HBL) H\,2356$-$309 ($z = 0.165$)
was first identified at optical frequencies by \citet{schwartz:1989a}.
Its first detection in the X-ray band was done by
satellite experiment UHURU \citep{forman:1978a}.
Later observations with BeppoSAX established it as an
\emph{extreme synchrotron blazar} \citep{costamante:2001a}.
%
\mbox{H\,2356$-$309} was discovered in the VHE gamma-ray regime by H.E.S.S.
after observations in 2004 \citep{2006A&A...455..461A}.
Simultaneous observations were carried out with RXTE in X-rays on 11th of November 2004,
with the Nan\c{c}ay decimetric radio telescope (NRT) between June and October 2004
and with ROTSE in the optical, covering the whole 2004 H.E.S.S. observation campaign.
With these data for the first time, an SED comprising simultaneous radio, optical, X-ray and VHE
measurements could be made.
The resulting broad-band SED from this observation campaign (Figure\,\ref{h2356_first_sed})
can be well described by a single-zone synchrotron self Compton (SSC) model, using a reasonable set of parameters.
%
A later multi-frequency campaign \citep[][]{2010A&A...516A..56H} with
H.E.S.S., NRT, ATOM, and XMM-Newton,
yielded similar results. The object still showed
little variability (factor 2) and was in a historically low X-ray-level.
A similar case of a broad-band SED compatible with a single-zone SSC model
is given by multi-frequency observations of the extreme synchrotron blazar 1ES\,1101-232 \citep{2007A&A...470..475A}.
Due to their comparatively large redshift, the results obtained from H\,2356$-$309
and 1ES\,1101$-$232 ($z = 0.186$) lead to strong constraints on the density of the EBL \citep{2006Natur.440.1018A}.

\subsection{PKS\,2155$-$304}
The high frequency peaked BL Lac object PKS\,2155$-$304 ($z=0.117$) is one of the best studied blazars in the southern hemisphere.
After first being detected in the X-ray band \citep{1978MNRAS.182..661C,1979ApJ...234..810G,1979ApJ...229L..53S},
it was detected in the VHE regime by the {\it University of Durham Mark 6 Telescope} in 1996 and 1997
\citep{1999ApJ...513..161C}.
H.E.S.S. could confirm these findings in early observations before the completion of the experiment
in 2002 and 2003 \citep{2005A&A...430..865A}.
In a subsequent multi-frequency campaign in 2003 \citep{2005A&A...442..895A},
the object was observed in a hisorically low flux level in X-rays.
Furthermore, the weak variability in the VHE H.E.S.S. data (smaller than factor 2.5) and the notable
fact that PKS\,2155$-$304 was detectable within each observation night by H.E.S.S.
indicated that the object was observed in a quiescent flux state.
Simultaneous multi-frequency observations in 2008
\citep[][H.E.S.S., Fermi, RXTE, and ATOM]{2009ApJ...696L.150A} yielded a clear optical/VHE
correlation and evidence for a correlation of the X-ray flux with the spectral index in the
high energy regime (Fermi data, E$>$100\,MeV). A behaviour that is difficult to explain in
the framework of SSC models.

The most notable events concerning this source, however, did happen in two exceptionally bright
VHE outbursts in 2006 \citep{2007ApJ...664L..71A,2009A&A...502..749A}.
In the first observation night covered by H.E.S.S., variability could be detected at the minute time-scale (3\,min bins).
Assuming a size of the emission region of the order of the Schwarzschild radius,
causality requires a Doppler factor of more than 100 in order to accomodate the observed
fast variability.
%
As shown by \citet{2008bves.confE..16D}, the data show log-normal flux variability in the VHE regime
\citep[also seen in Mrk\,421,][]{2010A&A...524A..48T}.
Such a variability pattern 
is indicative of a multiplicative process,
implying a connection of the VHE emission to accretion disk activity \citep[see][and references therein]{2009A&A...503..797G}.
Furthermore, a log-normal flux behaviour is incompatible with an additive process such as
a multi-zone IC emission scenario.

As opposed to the first observation night,
the second H.E.S.S. observation night \citep{2009A&A...502..749A} was also covered by observations by Chandra
(and partly by RXTE and Swift) in X-rays,
and the Bronberg optical observatory.
A clear VHE/X-ray correlation was found,
without significant time-lags between both frequency bands.
The main variability structure and sub-structures of the VHE lightcurves are matched in X-rays.
However, rapid sub-flares observed in the VHE band are not found in the Chandra data.
Huge differences in the variability amplitudes of \mbox{$\sim$10}/2/1.15 were observed between
the VHE/X-ray/optical bands.
The ratio of VHE (inverse Compton) to X-ray (synchrotron) flux was at a maximum of 8 during
the decaying phase of the flare, and reached a minimum at the end of the observation night.
Such a high and rapidly varying \emph{Compton dominance} was never observed before.

Spectral analysis of the H.E.S.S. data show a curved spectrum with a stronger curvature at high flux states,
indicating that the observed curvature is at least partly intrinsic.
Overall, the spectral index shows a harder-when-brighter relationship in the VHE regime.
While such a dependency is also seen in the simultaneous Chandra data, the partly sampled
X-ray flare starting after the H.E.S.S. exposure shows a softer-when-brighter relation.
%

\begin{figure}[t!]
\resizebox{\hsize}{!}{\includegraphics[clip=true]{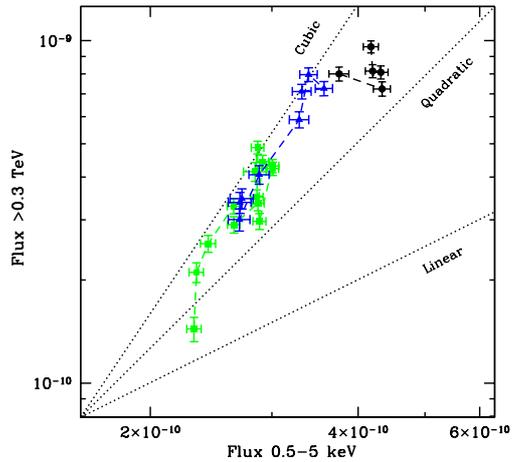}}
\caption{\footnotesize
Correlation between X-ray and VHE flux during 3 different states of the second exceptional flare (MJD53946):
maximum (black), decaying phase (blue), and
broad sub-flare (green). Mostly, a cubic relationship between the VHE
and X-ray bands is observed.
}
\label{pks2155_cubic}
\end{figure}

Remarkably, the gamma-ray flux traces the variations of the X-ray flux much stronger
than quadratically (Figure\,\ref{pks2155_cubic}). Instead, a cubic relöation is found.
As discussed in \citet{2009A&A...502..749A}, such a super-quadratic relationship
is challenging common scenarios of VHE emission in blazars.

\subsection{M\,87}
The giant radio galaxy M\,87 ($z = 0.0043$) is located at the center of the Virgo cluster of galaxies,
and hosts a central black hole of ($3.2\,\pm\,0.9$)$\cdot$10$^9$\,M$_\odot$.
As opposed to blazars, the jet of M\,87 is orientated at an angle of $\approx$\,30$^\circ$
relative to the line of sight. The possibility of multi-frequency observations with radio instruments
that are able to resolve sub-structures of its jet make M\,87 a very interesting object for studying the location of
the VHE emission.
After first indications for a VHE signal in HEGRA data \citep{2003A&A...403L...1A},
H.E.S.S. observations between 2003 and 2006 could establish M\,87 as the first VHE
emitting radio galaxy \citep{2006Sci...314.1424A}.
The spectrum seen by H.E.S.S. is well described by a pure powerlaw.
The hardness (spectral index $2.22\,\pm\,0.15$ in 2005) of the observed spectrum extending beyond 10\,TeV
and the lack of any absorption feature due to pair production
imply a low central infra-red (IR) luminosity at 0.1\,eV (10\,$\mu$m),
as could be expected in the case of an advection-dominated accretion disk.
It also excludes significant contributions
of the IR radiation fields as external seed photon population.
While the position of the VHE excess rules out the radio lobes as potential site for the gamma-ray production,
the resolution of the H.E.S.S. Cherenkov telescopes does not allow to resolve structures
smaller than the large scale lobe structure of M\,87.
However, the unexpectedly fast variability observed in H.E.S.S. data from 2005 allowed to strongly constrain the size of the
emission region.
Several potential sites and mechanisms for VHE gamma-ray production could thus be excluded:
the elliptical host galaxy,
dark matter, the kpc-scale jet, and the brightest knot (knot A) in the jet \citep[see][for discussion]{2006Sci...314.1424A}.
Remarkably, simultaneous observations carried out with the {\it Chandra} satellite
showed that the highest VHE flux state observed by H.E.S.S. coincided with a maximum X-ray flux from the HST-1 hotspot.

\begin{figure*}[t!]
\resizebox{\hsize}{!}{\includegraphics[clip=true,angle=270]{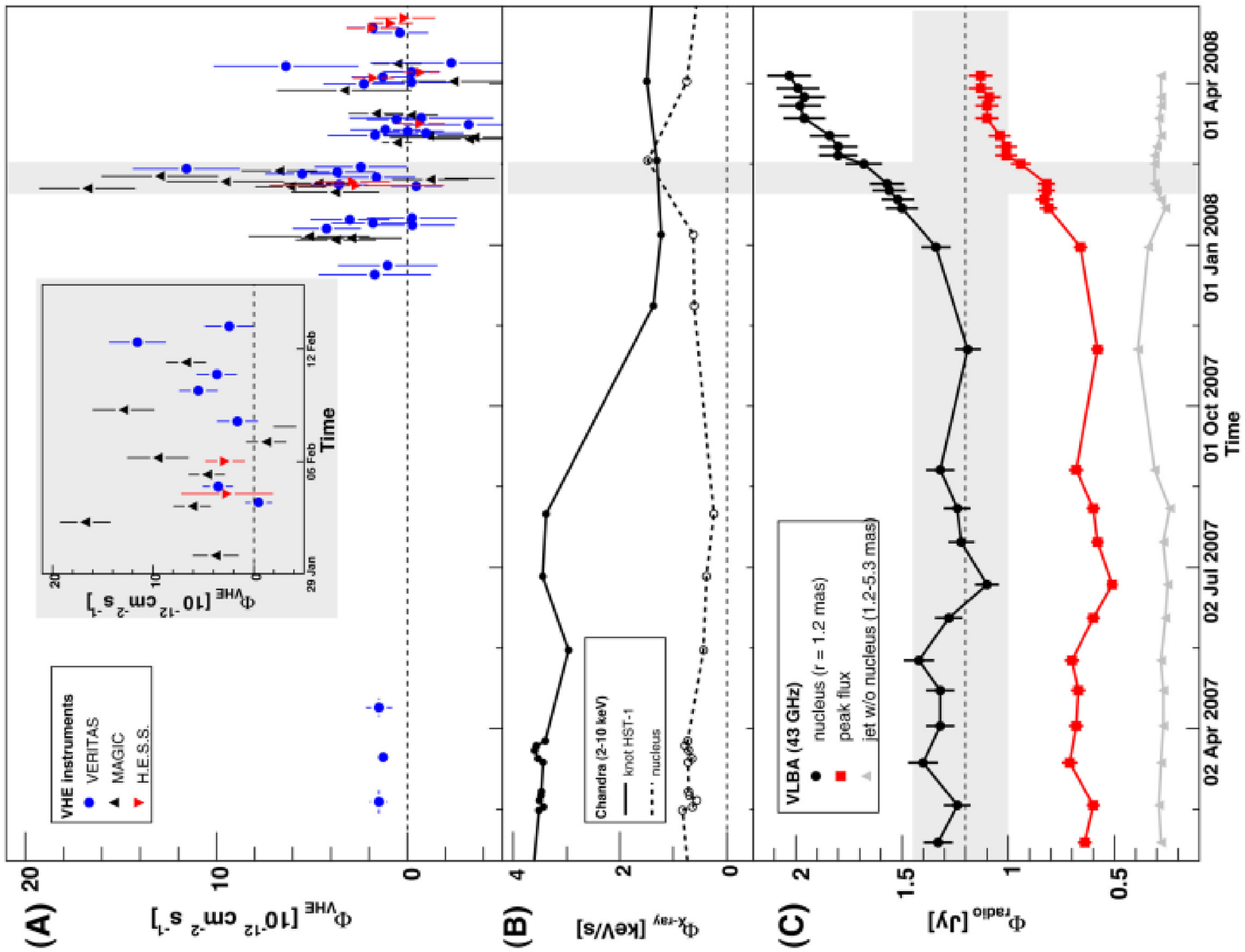}}
\caption{\footnotesize
This figure was taken from \emph{Science}, Vol. 325, p444 \citep[from][]{2009Sci...325..444A}.
Light curves for M\,87 (2007--2008). (A) nightly VHE gamma-ray fluxes (E$>$0.35 TeV).
(B) Chandra X-ray data (2 to 10 keV) of the nucleus and the knot HST-1 \citep{2009NJPh...11l3022H}.
(C) Flux densities from the 43-GHz VLBA observations of the nucleus,
the peak flux, and the flux integrated along the jet. For details, see \citet[][]{2009Sci...325..444A}.
}
\label{m87_radio_imaging}
\end{figure*}

In order to bypass the poor angular resolution of IACTs, a multi-frequency campaign was carried out
with participation of VERITAS, MAGIC, and H.E.S.S.,
together with the Very Long Baseline Array (VLBA) at 43\,GHz with a resolution of 0.21\,$\times$\,0.43\,mas,
and Chandra in the X-ray band \citep{2009Sci...325..444A}.
As opposed to the previous findings by H.E.S.S., the VHE maximum observed by all IACTs in 2008
occured during a minimum of X-ray activity from HST-1
(see Figure\,\ref{m87_radio_imaging}).
Instead, the VLBA data shows that the VHE flare occurs during a rise of activity in the nucleus.
These findings are an indication for the VHE emission to be taking place in the jet collimation
region \citep[see][]{2009Sci...325..444A}.
Recent data on M\,87 show an even more intriguing multi-frequency picture \citep[see e.g.][]{raue:extragalacticsky}.

%



\section{Summary}
In the past decade,
observations with H.E.S.S. and other instruments in other wavelength bands
have produced pieces of the puzzles of VHE emission from AGN
(See open questions, Introduction, page~\pageref{tluczykont_introduction}).

A log-normal flux pattern (multiplicative process) found in the VHE data of PKS\,2155-304
(and in combined data on Mrk\,421) might hint at a connection of the VHE emission to accretion disk activity,
and are contradictory to an additive multi-zone IC scenario.
During the second exceptional flare of PKS\,2155-304 observed by H.E.S.S.,
a cubic relationship was found between the X-ray flux and the VHE flux, challenging
common models of IC emission.
The unprecedentedly fast variability (minute time-scale)
seen from this source strongly constrains the size of the VHE
emission region.
A comparatively fast variability of the order of days was also found in
H.E.S.S. data on the radio galaxy M\,87, ruling out several hypotheses for
the location of the VHE emission region.
In a VHE/X-ray/radio multi-frequency campaign, 
evidence was found for a connection of the VHE emission from M\,87
to the jet collimation region.

\begin{acknowledgements}
The support of the Namibian authorities and of the University of Namibia in facilitating the
construction and operation of H.E.S.S. is gratefully acknowledged, as is the support by the
German Ministry for Education and Research (BMBF), the Max Planck Society, the French Ministry
for Research, the CNRS-IN2P3 and the Astroparticle Interdisciplinary Programme of the CNRS,
the U.K. Science and Technology Facilities Council (STFC), the IPNP of the Charles University,
the Polish Ministry of Science and Higher Education, the South African Department of Science
and Technology and National Research Foundation, and by the University of Namibia. We
appreciate the excellent work of the technical support staff in Berlin, Durham, Hamburg, Heidelberg,
Palaiseau, Paris, Saclay, and in Namibia in the construction and operation of the equipment.
The author thanks the organizers for their kind invitation to this inspiring workshop.
\end{acknowledgements}

\bibliographystyle{aa}

\bigskip
\bigskip
\noindent {\bf DISCUSSION}

\bigskip
\noindent {\bf SERGIO COLAFRANCESCO:} It seems that the radio high-resolution observations are crucial for the
location of the VHE emission. Do you also have radio polarization observations to better identify the radio
blob responsible for the VHE emission ?

\bigskip
\noindent {\bf MARTIN TLUCZYKONT:} We don't have results from polarization measurements yet.
P.S.: during the session, the answer referred to results instead of existence of data.
Data exist and analysis is in progress.

\bigskip
\noindent {\bf JIM BEALL:} Can you comment on the size scale where the VHE / radio emission occurs ?

\bigskip
\noindent {\bf MARTIN TLUCZYKONT:}
From causality arguments alone, the fast variability observed from M\,87 implies an absolute \emph{size} of the
emission region of the order of 3 Schwarzschild radii of the central black hole ($3\,\times\,10^{15}$\,cm).
The absolute \emph{position} of the VHE emission can be derived to be within the jet collimation region,
considering the simultaneity of the VHE flare with the radio emission from the nucleus in 2008.

\end{document}